\documentclass[10pt,letterpaper]{article}
\usepackage{opex3}
\usepackage{amsmath}

\begin{document}

\title{Highly efficient heralding of entangled single photons*}

\author{Sven Ramelow$^\dagger$,$^{1,2}$ Alexandra Mech$^\dagger$,$^{1,2}$ Marissa Giustina$^\dagger$,$^{1,2}$ Simon Gr\"oblacher,$^{3,4}$ Witlef Wieczorek,$^{4}$ Adriana Lita,$^{5}$ Brice Calkins,$^{5}$ Thomas Gerrits,$^{5}$ Sae Woo Nam,$^{5}$ Anton Zeilinger,$^{1,2,4}$ and Rupert Ursin$^{1}$}

\address{$^1$Institute for Quantum Optics and Quantum Information, Austrian Academy of Sciences, Boltzmanngasse 3, Vienna, Austria \\
$^2$Quantum Optics, Quantum Nanophysics, Quantum Information, University of Vienna, Faculty of Physics, Boltzmanngasse 5, Vienna, Austria \\
$^3$current address: Institute for Quantum Information and Matter, California Institute of Technology, 1200 E. California Blvd., Pasadena, CA 91125, USA\\
$^4$Vienna Center for Quantum Science and Technology, Faculty of Physics University of Vienna, Boltzmanngasse 5, Vienna, Austria\\
$^5$NIST, Boulder, CO, USA\\
$^\dagger$ These authors contributed equally to this work\\}

\email{sven.ramelow@univie.ac.at, marissa.giustina@univie.ac.at} 



\begin{abstract*} Single photons are an important prerequisite for a broad spectrum of quantum optical applications. We experimentally demonstrate a heralded single-photon source based on spontaneous parametric down-conversion in collinear bulk optics, and fiber-coupled bolometric transition-edge sensors. Without correcting for background, losses, or detection inefficiencies, we measure an overall heralding efficiency of 83~\%.  By violating a Bell inequality, we confirm the single-photon character and high-quality entanglement of our heralded single photons which, in combination with the high heralding efficiency, are a necessary ingredient for advanced quantum communication protocols such as one-sided device-independent quantum key distribution.\\

*Partial contribution of NIST, an agency of the U.S. government, not subject to copyright
\end{abstract*}




\section{Introduction}

The controlled and deterministic generation of single-photon states and correlated pairs remains a challenge particularly crucial to a wide variety of emerging optical quantum technologies including metrology \cite{rarity_absolute_1987, kwiat_absolute_1994, migdall_absolute_1995}, quantum communication \cite{gisin_quantum_2002, scarani_security_2009, MZ_rev_2012} and optical quantum computing \cite{kok_linear_2007, ladd_quantum_2010}, to name just a few. Although they are not inherently deterministic, highly efficient heralded single-photon sources are relevant to this problem. By combining such a heralded source with a photon memory that can store and release photons in a controlled way \cite{pittman_single_2002, jeffrey_towards_2004} or by multiplexing several heralded sources and using feed-forward and fast switching to select a channel that contains a single photon \cite{migdall_tailoring_2002,shapiro_demand_2007}, it is possible to construct an on-demand single- (and by extension, multi-) photon source. Such a source could be an important ingredient for post-selection-free multi-photon one-way quantum computation \cite{PhysRevA.68.022312}. Even without these extensions, a highly efficient heralded photon source would be valuable. For example, the intrinsically secure one-sided device-independent quantum key distribution protocol requires sources of entangled photons with heralding efficiencies (including detection) of at least 66 \% \cite{PhysRevA.85.010301}, which have not yet been demonstrated \cite{smith_conclusive_2012}. 
In addition, such high coupling efficiency marks an important step toward a loophole-free Bell test as it is relevant not only to the fair-sampling loophole but also to the freedom-of-choice loophole \cite{private_comm_FOC, AZ_FOC}.
Furthermore, any source that heralds the arrival of a known photon number and energy is also useful for coincidence-based detector calibration \cite{rarity_absolute_1987, kwiat_absolute_1994, migdall_absolute_1995, avella_11}, which promises to overcome the precision limitations of power and attenuation measurements that presently dominate the calibration process for single-photon detectors.

An ideal heralded photon source should provide a heralding signal that indicates the presence of exactly one photon, preferably in a fiber. Over the past decades, spontaneous parametric down-conversion (SPDC) has proven to be a robust, well-understood, and reliable method for generating time-correlated photon pairs that may be split into two spatial modes. SPDC is a promising candidate for high-quality heralded single-photon sources where the detection of a photon in one mode (the heralding or idler mode) indicates a photon in the other (signal mode). Moreover, producing polarization entanglement based on SPDC has been demonstrated with high quality and flexibility, which suggests that such heralding sources are readily extendible to the quantum applications listed above \cite{kwiat_new_1995, kim_phase-stable_2006, fedrizzi_wavelength-tunable_2007}. Although fiber coupling demands additional precision in the construction of the source, it substantially improves versatility; furthermore the mode selection achieved by the fiber can enhance the heralding efficiency.

\begin{figure}
\centering
\includegraphics[scale= 0.25]{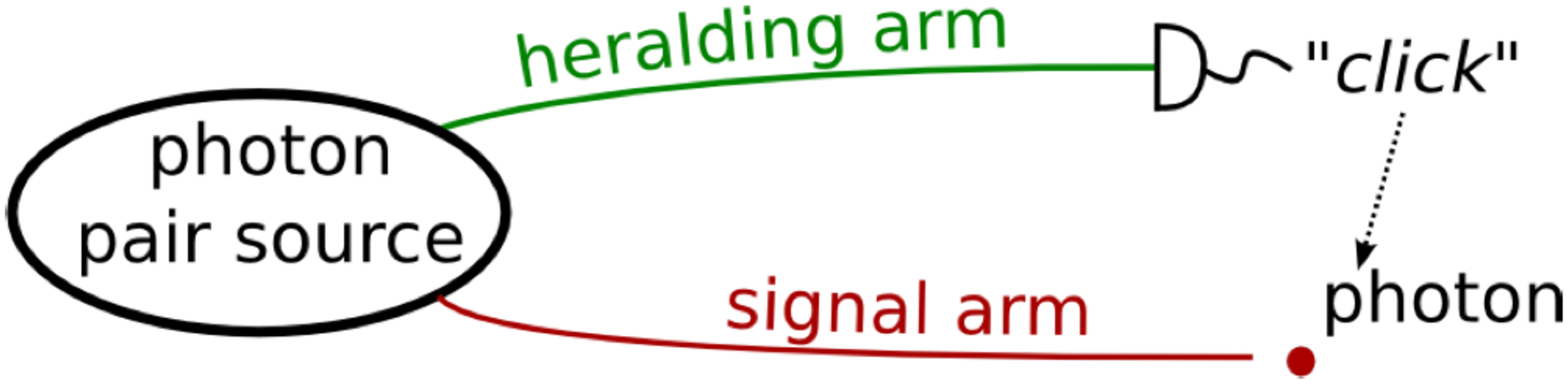}
\caption{Heralded single-photon source based on correlated photon pairs. Such sources are a prerequisite to a multitude of quantum optical experiments. In an ideal single-photon source, a photon detected in the heralding arm indicates a partner photon in the signal arm.}
\end{figure}

In practice, no source is truly ideal and any source may be subject to a ``missing-photon error," such that a heralding signal is issued but no photon is present in the signal mode. This may result from background in the heralding signal or photon loss in the source or signal mode, for instance from imperfect optical elements or fiber coupling. Note that to ``directly observe" a given heralding efficiency value, that is, to measure it without correction for detector inefficiencies, it is necessary to utilize a detector which itself is at least as efficient as this value.

Due to missing-photon errors and the low efficiency of the industry-standard silicon avalanche photodiodes (APDs), the highest reported directly observed heralding values (without correction for detector inefficiencies) have until recently been in the range between $30~\%$ and $50~\%$ for wavelengths between 600 nm and 850 nm \cite{trojek_collinear_2008, soller_high-performance_2011, wittmann_loopholefree_2012}. However, the recent advent of superconducting bolometric detectors suggests the impending reality of near-unity heralding efficiency, and, utilizing this technology, total heralding efficiencies of up to $62~\%$ have already been observed \cite{smith_conclusive_2012}. Here, we combine the nearly-perfect detection efficiency offered by transition-edge sensors (TES) \cite{lita_counting_2008} with the ultra-high coupling efficiency of our fiber-coupled photon pair source based on SPDC in a bulk crystal; we achieve a source in which up to $83~\%$ heralding efficiency has been observed. In addition, when using polarization-entangled photon pairs we record the only slightly reduced heralding value of $80~\%$. (Note that the efficiencies reported here have been measured \emph{directly}, without correction for dark counts, accidental counts, inefficient detection, or known optical losses.) To our knowledge, these values are the highest directly observed heralding efficiencies. We analyze the origin of the remaining losses in our system; this analysis confirms the nearly-perfect efficiency of the TES detectors and indicates that it may be feasible to observe heralding values close to $100~\%$ with the presented technology based on bulk-crystal down-conversion and TES detectors.

\section{Experiment}

\begin{figure}
\centering
\includegraphics[scale=0.45]{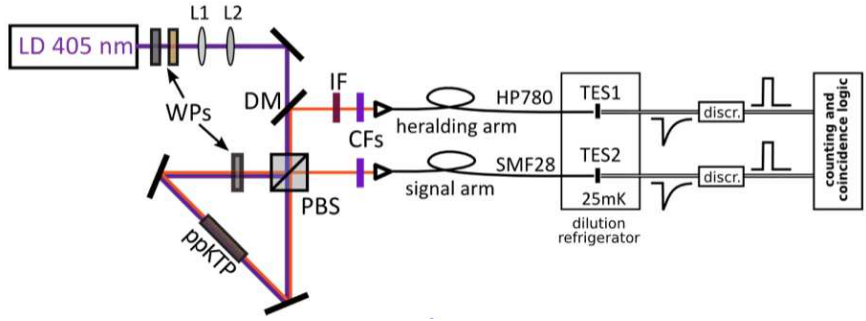}
\caption{Experimental setup: The photon pair source is based on a  10 mm long ppKTP crystal pumped by a 405 nm diode laser in a Sagnac configuration \cite{fedrizzi_wavelength-tunable_2007} with a polarizing beam splitter (PBS). Waveplates (WPs) are used to tune the pump polarization. The pump beam is carefully shaped and focussed by two lenses (L1, L2) and split from the down-converted photon with a dichroic mirror (DM).  Cut-off filters (CFs) are used to filter out the remaining 405 nm light and a narrowband interference filter (IF) in the heralding arm further suppresses any photons not originating from the down-conversion. The photon pairs are coupled into optical fibers (HP780, SMF28) that carry them into the dilution refrigerator where they are directly coupled to the TES detectors with their SQUID amplifiers (TES1, TES2) which are held at around 25 mK. The TES output signals are discriminated using threshold discrimination and are counted and analyzed by our coincidence electronics.}
\end{figure}

Our source of (entangled) photon pairs (see Fig.2) is based on SPDC in a periodically poled potassium titanyl phosphate (ppKTP) crystal with a poling period of around $10~ \mu m$ for the type\nobreakdash-II quasi-phase-matched creation of photon pairs at 810 nm with a 405 nm pump diode laser -- the design is described in detail in \cite{kim_phase-stable_2006,fedrizzi_wavelength-tunable_2007}. The crystal can be pumped bi-directionally in a Sagnac-type configuration to produce polarization entanglement. Pumping the crystal in only one direction creates a polarization product state. The emitted photon pairs are split, with each photon entering one of two (separate) single mode (SM) fibers.

Any photon detected in the heralding arm announces (heralds) the presence of a photon in the signal arm. To reach a high heralding efficiency it is crucial to detect in the heralding arm only photons from the SPDC process, while simultaneously minimizing losses in the signal arm. This is achieved in our setup by a number of steps. Tight spectral filtering of the heralding arm, accomplished with a cut-off filter to block pump photons (longpass filter with cutoff around 650 nm) and a 1 nm bandpass filter centered at the down-converted wavelength, ensures that only photons directly emitted by the SPDC process (intrinsic bandwidth $\approx 0.5 $ nm FWHM) are coupled to the heralding arm. The inevitable loss introduced by this spectral filtering does not reduce the heralding efficiency, which depends on the transmission in the signal arm. The signal arm is filtered only with a cut-off filter (as used in the heralding arm) to suppress the pump light, which introduced a loss of around 2 \% for 810 nm. In addition, we optimize the focusing parameters and spatial shaping of the pump beam and heralding arm to maximize the heralding efficiency \cite{fedrizzi_wavelength-tunable_2007, bennink_optimal_2010}. By coupling to a standard single-mode fiber (HP780) in the heralding arm, but to a standard telecom fiber (SMF-28) -- which carries two TEM modes of 810 nm light -- in the signal arm, the heralding efficiency can be further increased. The fiber tips were anti-reflection (AR) coated for 810 nm to minimize reflection losses in the fiber coupling.

Photon detection is accomplished in our experiment with transition-edge sensors (TES), which in recent years have attracted considerable attention as highly efficient single-photon counters \cite{lita_counting_2008, Fukuda_11}. For detecting photons in the visible and near-infrared regime, a $25~ \mu m$ square of tungsten thin film, cooled to well within the superconducting state and voltage-biased to well within the superconducting transition, serves as both absorber and thermometer in this bolometric-style detector \cite{irwin_1995}. Any photon absorbed by the tungsten will heat it and manifest an increase in resistance, which in turn yields a proportional current drop though the voltage-biased device on the order of 50~nA. Then the heat dissipates through a weak thermal link to the base temperature, and the detector returns to its original resistance. 
Although the TES film itself has a thickness of only 20~nm, embedding the TES in a wavelength-specific optical cavity yields detectors with peak efficiencies of at least 95~\% for the selected wavelength \cite{lita_counting_2008, Miller:11}. Note that this value includes losses in coupling from an AR-coated SMF-28 fiber to the TES chip, which is accomplished by a packaging process detailed in \cite{Miller:11}.

The TES is operated in series with an input coil that is inductively coupled to a superconducting quantum interference device (SQUID) for readout \cite{SQUIDs}. The signal spike from an incident photon enters the SQUID as a changing flux and may be read out as a voltage. In the relevant bandwidth of our electrical measurements (up to 1 MHz) the SQUID's input-referred current noise is less than 25~\% of that of the TES output current, so the TES itself dominates the noise of the system. Photons were distilled from the analog electrical output signal according to the following procedure. Individual photon spikes were identified and converted to TTL pulses using a leading-edge discriminator. We set the threshold of this discriminator to a value that registered a reasonable count rate of 810 nm photons but also minimized the dark count rate when the source was blocked. Although TES detectors have no intrinsic dark counts, and only a real energy signal will create a current pulse, a non-zero background level may be registered by the presence of background light in the experimental setup \cite{rosenberg_noise-free_2005}. Even thermal blackbody radiation or infrared photons may be seen by a TES optimized for use in the visible regime, and if the threshold level is set too close to zero, such a thresholding counting method may lead to increased background counts.

\begin{figure}
\centering
\includegraphics[scale=0.5]{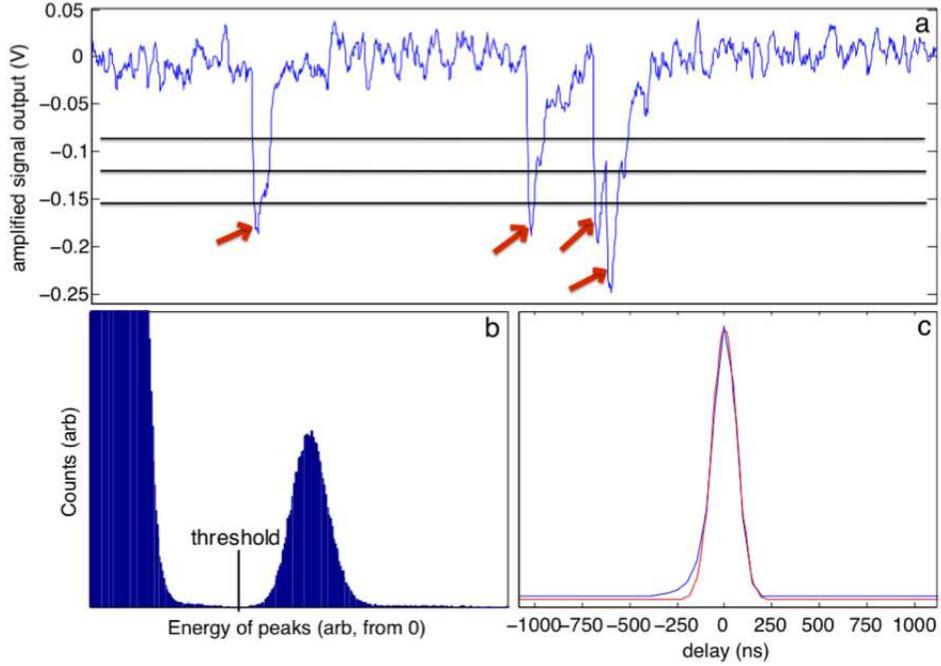}
\caption{Photon signals and processed data from transition-edge sensor single-photon detectors. (a) A typical signal from a detector with four photons and different possible thresholds indicated. The top threshold detects only three of the four photons, the middle threshold counts five (one from the wiggle in the recovering edge) and the bottom threshold detects the correct four photons. (b) A ``pulse height distribution," indicating how clearly it is possible to separate the photon signals from the noise by thresholding. (c) Coincidence count rates vs. delay between the two channels. The actual data is plotted in blue with a Gaussian fit in red. Asymmetry is attributed to uneven detector jitter.}
\end{figure}

To avoid re-triggering and thereby mistakenly counting a non-existent second photon in the noise of the recovering edge of the first, we implemented a ``deadtime" by using TTL pulses sufficiently long to ``re-arm" the discriminator only after the signal's recovery. Thereafter we counted coincidences using an analog logic module that registered a coincidence for each overlap of greater than 3 ns between the TTL pulses of the two channels. Thus the effective coincidence window is defined as the sum of the TTL pulse lengths for the two channels, which in our case was $1.05~\mu$s: 1 $\mu$s for the heralding arm and 0.05 $\mu$s for the signal arm. Each coincidence was represented by yet another TTL pulse from the logic module, and all three TTL channels were counted with a standard counter connected to a PC, which allowed us to monitor the heralding efficiency in real time.

As an alternative to analog discriminators and logic modules, we also digitized and recorded data for post-processing using a data acquisition board. Post-processing facilitates more complicated counting algorithms and finer control over the coincidence window. These algorithms also enable the recovery of photons lost by the analog counting method. More information on post-processing will be detailed in a subsequent paper \cite{MG_in_preparation}; more information on the correction of so-called ``accidental coincidences" may be found in the appendix.

\section{Results}
We tuned the pump power to a level that produced a photon rate suitable for the detectors and pumped the source in only one direction (creating a polarization product state). With the source in this condition, we measured the singles and coincidences for 100 seconds with our analog electronics and digitized 40 seconds of data for post-processing. The results are summarized in Table 1. The raw ratio between heralding counts and directly observed coincidences is $83.1~\% \pm 0.2~\%$. This represents an unprecedented value for uncorrected heralding efficiency.

\begin{table}
\centering
\begin{tabular}{|c|c|c|c|}
\hline \rule[-2ex]{0pt}{5.5ex}  & signal arm & heralding arm & coincidences \\ 
\hline \rule[-2ex]{0pt}{5.5ex} analog-processed counts & 46855 $\pm$ 22 s$^{-1}$ & 6525 $\pm$ 8 s$^{-1}$ & 5419 $\pm$ 7 s$^{-1}$ \\ 
\hline \rule[-2ex]{0pt}{5.5ex} analog arm efficiency & 83.0\% $\pm$ 0.2\% & 11.57\% $\pm$ 0.02\% &  \\ 
\hline \rule[-2ex]{0pt}{5.5ex} acc.-corrected efficiency & 82.0\% $\pm$ 0.3\% & 11.39\% $\pm$ 0.03\% &  \\ 
\hline \rule[-2ex]{0pt}{5.5ex} post-processed counts & 49882 $\pm$ 35 s$^{-1}$ &7696 $\pm$ 14 s$^{-1}$ &  6303 $\pm$ 13 s$^{-1}$ \\ 
\hline 
\end{tabular}  
\caption{Tabulated experimental results from both the analog electronics counting method and the post-processing for 100 seconds and 40 seconds of data respectively. The lower efficiency in the second arm is a consequence of the higher loss caused by the limited transmission efficiency of the narrow bandpass filter as well as a high rate of background photons in the signal arm not rejected by the cut-off filters. The post-processing method can recover counts not registered by the analog method. The one standard deviation errors are determined by Poissonian counting statistics and error propagation.}
\end{table}

When determining the heralding efficiency as the ratio of the measured coincidence and single count rates, it is necessary to account for a systematic error known as ``accidental coincidences." An accidental coincidence occurs each time two photons that did not originate from the same pair are detected within a coincidence window and are thus counted as a coincidence. If left uncorrected, this effect would lead to an over-estimation of the actual heralding efficiency. Using the formulas explained in the appendix, we find a systematic error for the heralding efficiency of $1.0\pm 0.1~\%$ which leads to a corrected heralding efficiency of $82.0 \pm 0.3 ~\% $. Post-processing, which allows us to choose our coincidence window and includes accidental correction by a method similar to that described in the appendix, yields the heralding value of  81.9\% $\pm$ 0.2~\% for the same coincidence window of 1.05 $\mu$s, which agrees very well with the directly observed data. We also measured a combined jitter of approximately 155 ns FWHM for the two detectors, which determines an upper limit on the timing precision with which we herald our photons. Note that this could be further improved, without affecting the heralding efficiency, by replacing the heralding detector with a low-jitter detector.

The system detection efficiency of the TES is expected to be close to unity \cite{lita_counting_2008}. In our experiment, the following losses contribute to the reduction of the heralding efficiency from 100~\%. The estimated total optical losses in the source sum to around $6~\%$ \cite{fedrizzi_wavelength-tunable_2007}.  Additionally, there are losses due to fiber coupling; using standard silicon APDs to compare the heralding ratios between large core diameter fibers (multimode, 50~$\mu$m) and the SMF-28 fiber used in the experiment, we concluded that the loss introduced by the fiber coupling is around $10~\% $. Combining these estimations with the measured and accidental-corrected heralding efficiency of $82~\%$, we find that the system efficiency of the TES (including fiber splices, interface between fiber and detector, and quantum efficiency of the absorptive area of the TES) is with high certainty above $95\%$~\cite{lita_counting_2008}. This represents the first verification of the transition-edge sensor's near-unity detection efficiency using a method based on the quantum nature of light and thus differing from the standard approaches based on power measurements and calibrated attenuation.

Entanglement is a necessary ingredient in the system if one wants to prepare heralded single photons in a remotely chosen basis or utilize the high heralding efficiency for one-sided device-independent QKD \cite{smith_conclusive_2012}. To generate entanglement in our source, we pumped it in both directions to produce a nearly maximally-entangled state \cite{fedrizzi_wavelength-tunable_2007}. In this state, the directly-measured heralding efficiency decreased to $79.7\% \pm 0.2~\%$. We believe the reduction is mainly a consequence of imperfect overlap between the two pump directions when the source is pumped bi-directionally.
To verify a high degree of entanglement we tested a CHSH inequality \cite{CHSH}. For the necessary  polarization measurements we inserted plate polarizers with an additional loss of around $15~\%$ and measured the polarization correlations for all necessary setting combinations for the CHSH inequality, integrating for 10 s per setting. This resulted in a Bell parameter of $S = 2.51 \pm 0.01$ which is more than 50 standard deviations above the classical bound of 2 and shows a high fidelity of the entangled state. Note that the presence of entanglement confirms the single-photon nature of our source.

\section{Conclusion and Outlook}

In conclusion, using TES detectors we demonstrated a heralded single-photon source with an unprecedented high efficiency in bulk optics, achieving a single-photon heralding efficiency of 83~\% with no correction for background, detection efficiency or other losses. Moreover, it was possible to produce heralded entangled photons, useful for one-sided device-independent QKD, with only minimal decrease in efficiency.

To compensate for the systematic accidental coincidences which would otherwise lead to an overestimation of heralding efficiency, we developed an extended accidental correction model. This takes into account the very high coupling efficiencies, which are typically assumed to be small to justify neglecting several terms in the expression quantifying the expected accidentals. 

Moreover, we would like to point out that our source facilitated the use of heralded single photons to infer a system detection efficiency of over 95~\% for a TES detector. It is important to note that for a detector-calibration method such as this, which is based on correlated photons, the accuracy of the measurement improves with the heralding ratio.

Finally, we note that while very promising fiber-based realizations of heralded single-photon sources have already been demonstrated \cite{soller_high-performance_2011}, our results indicate that with further optimization of losses and focusing conditions, it should be possible to reach near-unity heralding efficiencies in a bulk-optics configuration.

\section{Acknowledgement}
The authors acknowledge the help of J\"orn Beyer and Marco Schmidt at the Physikalisch-Technische Bundesanstalt in Berlin, Germany for providing SQUID amplifiers and assistance with the TES-SQUID system setup. In addition, we thank Markus Aspelmeyer for the use of the dilution refrigerator.

This work was supported by the Austrian Science Foundation (FWF) under projects SFB F4008 and CoQuS, the grant Q-ESSENCE (no. 248095), and the John Templeton Foundation. SG acknowledges support from the European commission through a Marie Curie fellowship. WW
acknowledges support from the Alexander von Humboldt Foundation and is a recipient of a Marie Curie Fellowship of the European commission. This work was also supported by the NIST Quantum Information Science Initiative (QISI).

\newpage

\section{Appendix}

\subsection{Accidental effects}

For determining the systematic effects of accidental coincidences a careful analysis is required. Accidental coincidences occur when two photons that are not from the same pair are detected and counted as a real coincidence. Their rate depends on the length of the coincidence window $\tau_{w}$, the rate of produced pairs $R_0$, and the two total arm efficiencies $\eta_1$ and $\eta_2$.

Assuming dark counts are negligible, the two singles rates (total rate of detected clicks per detector) for each arm $S_1$ and $S_2$ are given by:
\begin{eqnarray}
S_1=R_0\eta_1 \\
S_2=R_0\eta_2.
\end{eqnarray}

If the detectors have a dead-time $\tau_{d}$ (time interval after a detection event in which the detectors are blind) this will create a saturation effect which, for dead-times much smaller then the inverse detection rates, leads to singles rates given by the following:
\begin{eqnarray}
S_1=R_0\eta_1 (1- S_1 \tau_{d} )\\
S_2=R_0\eta_2 (1- S_2 \tau_{d} ).
\end{eqnarray}

These can be derived by the following argument: for any detected photon (detected at a rate $S$) there is a probability of $S\tau_{d}$ that a second photon appears within $\tau_{d}$ after the detection of the first. Such a photon would be lost because of the blind detector. Therefore, the rate of lost detections is $S^2\tau_{d}$, or equivalently a correction factor of $(1-S\tau_{d})$ must be used. Note that, in general, there can be different dead-times for the two detectors.

A coincidence is defined as the detection of two photons, one in each arm, separated by a time difference of less than half the given coincidence window, i.e., with a time difference between $-\tau_{w}/2$ and $+\tau_{w}/2$ (resulting in a full window of $\tau_{w}$). With negligible accidental coincidences (e.g., in the limit of very small pair-creation probabilities per coincidence window) one would detect the following (unmodified) rate of coincidences $CC_0$:
\begin{equation}
CC_0= R_0\eta_1 \eta_2.
\end{equation}

However, as described above, sometimes two (unrelated) pairs are \textit{accidentally} created so close to each other that photons from different pairs may be detected as a coincidence. These are generally called \textit{accidental coincidences}. Note that in most cw photon-pair source implementations, the coherence time of the produced photons will be much shorter (order of picoseconds) than the coincidence window. Thus for cw down-conversion, higher order contributions (genuine multi-pair emission) are typically negligible. 

There are now two possibilities by which accidental coincidences may occur. First, the first photon of the accidental coincidence is detected in arm 1, while its partner photon in arm 2 is lost -- this happens with a rate of $R_0\eta_1 (1-\eta_2)$. In order to lead to an accidental coincidence a second pair needs to be created within half the coincidence window after the detection of the first photon, which happens with a probability of $R_0\cdot\tau_w/2$. To cause a coincidence, the photon in arm 2 must be detected, which happens with efficiency $\eta_2$. Importantly, the rate of coincidences is increased only if this second pair would not have otherwise been detected as a coincidence on its own, meaning that its partner photon in arm 1 must not have been detected. The probability for this is given by $(1-\eta_1)$. Collecting all the terms, the rate $R_{10}$ by which the detected coincidences are increased for this case is given by:
\begin{equation}
R_{10} = \frac{1}{2}R_0^2\tau_{w}\eta_1 (1-\eta_2)\eta_2(1-\eta_1).
\end{equation}
In the same way one can derive the rate increase $R_{01}$ caused by the second possibility -- where the first photon of the accidental coincidence is detected in arm 2, while its partner is not detected in arm 1, and simultaneously a photon in arm 1 is detected from a second pair that is created within half a coincidence window after the detection of the first photon and would not have been detected as a coincidence on its own:
\begin{equation}
R_{01} = \frac{1}{2}R_0^2\tau_{w}\eta_2 (1-\eta_1)\eta_1(1-\eta_2).
\end{equation}

$R_{10}$ and $R_{01}$ are actually the same, as one would expect, since only the time ordering of the respective events is reversed and this does not change their probability. However, dividing the cases into these two distinct possibilities makes the logic of the argument easier to follow.

There is also an effect that reduces the number of detected coincidences. This is a saturation effect that also depends on how exactly the coincidence logic is technically implemented. A commonly used method is to create a pulse or bin with a length of half the coincidence window for each of the detector channels.  A coincidence is then counted for each overlap of pulses from the two different channels -- i.e., when the two detection events happen with a time difference between $-\tau_w/2$ and $+\tau_w/2$. When two pairs are created within a time span of $\tau_w/2$ and both photons of the first pair are detected (which happens with a rate of $R_0\eta_1\eta_2$), the second pair (which occurs with probability $R_0 \tau_w/2 $ and is detected with probability $\eta_1\eta_2$) cannot be detected as a coincidence anymore. This is similar to the dead-time effect for the singles rate. The rate of coincidences is therefore reduced by the number of events that would have been detected without this effect. This reduction of coincidences is given by:

\begin{equation}
R_{11} = - \frac{1}{2}\tau_{w} R_0^2\eta_1^ 2\eta_2^2.
\end{equation}

If the pulse or bin lengths for the two arms are different, the longer of the two ($\tau_{max}$) will be the effective dead-time instead of $\tau_{w}/2$. Importantly, if the intrinsic dead-time of the detectors is greater than $\tau_{max}$ or $\tau_{w}/2$ then $\tau_d$ must be used for $R_{11}$ instead of $\tau_{max}$ or $\tau_{w}/2$.

Taking now all three contributions ($R_{10}$, $R_{01}$ and $R_{11}$) into account, the rate of observed coincidences $CC=CC_0+R_{10}+R_{01}+R_{11}$ is given by:

\begin{eqnarray}
CC = R_0 \eta_{1} \eta_{2} &+ \tau_{w} R_0^2 \eta_1 \eta_2 (1-\eta_1)(1-\eta_2) \\
\nonumber &- \frac{1}{2} \tau_{w} R_0^2 \eta_1^2\eta_2^2,
\end{eqnarray}
or more compactly written:
\begin{equation}
CC = CC_0 (1+\tau_{w} R_0 (1-\eta_1)(1-\eta_2) -\frac{1}{2}\tau_{w}R_0\eta_1\eta_2),
\end{equation}
and if the pulse lengths differ between the two arms, 
\begin{equation}
CC = CC_0 (1+\tau_{w} R_0 (1-\eta_1)(1-\eta_2) -\tau_{max}R_0\eta_1\eta_2).
\end{equation}

The last equation and the two equations for the singles rates $S1$ and $S2$ form a set of three equations for the three unknown quantities  $\eta_1$, $\eta_2$ and $R_0$. These can be determined by solving this set of equations given the experimental parameters (the $\tau$'s) and measured rates $S1$ and $S2$ to yield the accidental (and dead-time) corrected values for $\eta_1$ and $\eta_2$. The general full solutions for these are rather long formulas, however these can be easily handled by mathematics software. 

Using the measured rates of $S1 = 46855.2~\textrm{s}^{-1}$, $S2 = 6525.0~\textrm{s}^{-1}$, and $CC = 5418.8~\textrm{s}^{-1}$, as well as $\tau_w=1.05~\mu s$ and $\tau_{max} =1~\mu s$, yields the accidental corrected value of $82.0\% \pm 0.3 \%$ for $\eta_1$.

\end{document}